\documentclass[twocolumn,a4paper]{article} 
\usepackage{amsmath,amssymb}
\usepackage{bm}
\usepackage{rev2002a,bp2312,ODE2008,z17}
\usepackage{name26}
\makeatletter\if@twocolumn\relax\else
\renewenvironment{multline}{\equation}{\endequation}\fi
\makeatother
\newcommand{\Ell}{\ell_{12}}%

\newcommand{\DeM}[1][\relax]{{#1\Delta}_{\mathrm{m}}}

\newcommand{\cosDeM}{c_{\mathrm{m}}}
\newcommand{\sinDeM}{s_{\mathrm{m}}}
\newcommand{\WK}{W_{\mathrm{kin}}}
\newcommand{\WP}{W_{\mathrm{pot}}}
\newcommand{\fd}{f_{\mathrm{d}}}
\newcommand{\omD}{\omega_{\mathrm{d}}}

\newcommand{\Deg}{{}^\circ}
\newcommand{\Hz}{\mathrm{Hz}}
\newcommand{\cm}{\mathrm{cm}}
\newcommand{\kg}{\mathrm{kg}}
\newcommand{\vort}{\bm{\omega}}
\usepackage{graphicx}
\graphicspath{{/Users/ooshida/talk/jps2603/}%
              {/Users/ooshida/Pictures/FigPS/}{.}}
\newcommand{\IncFig}[2][clip]{\includegraphics[clip,#1]{#2}}
\title{Experimental detection of energy transfer 
       into the anti\-phase mode in a branched double pendulum}
\author{Y.~Toda and 
        T.~Ooshida\thanks{Corresponding author: ooshida@tottori-u.ac.jp}}
\renewenvironment{abstract}{\begin{quotation}\noindent}{\end{quotation}\vspace{2em}}
\begin{document}
\twocolumn[
\maketitle
\begin{abstract}%
Multiple pendulum with branching
  is proposed as a convenient platform
  to study energy transfer between different modes.
The anti\-phase oscillation mode is localized
  to the ``child'' links,
  which makes it easy to prepare initial conditions
  without exciting the antiphase mode.
A manageable expression for the energy transfer
  is derived theoretically and evaluated with experimental data.

\end{abstract}]
\section{Introduction}

The arrangement of the pendulum system, 
  with which we performed experiments of energy transfer, 
  is motivated by studies of chaotic fluid motions 
  commonly referred to as turbulence \cite{Frisch.Book1995}.
Familiar flows of water and air
  are governed by the Navier--Stokes equation, 
  which can be written in the form of the vorticity equation,
\begin{equation}
  \left( \dd{}{t} + \mathbf{u}\cdot\nabla - \nu\nabla^2 \right) \vort
  = \vort\cdot\nabla\mathbf{u}
  \label{NS.vort}, 
\end{equation}
  where $\mathbf{u}$ is the velocity field (assumed to be solenoidal) 
  and $\vort = \nabla\times\mathbf{u}$ is the vorticity field.
The viscosity $\nu$ has negligibly small effect 
  as long as the velocity field remains smooth 
  at the human scale or above, 
  so the mechanical energy is conserved. 
However, 
  as the large scale flows 
  stretch and amplify smaller vortices \cite{Goto.RMPP8}, 
  the energy of large vortices is transferred to the smaller ones 
  and cascades down to even smaller eddies, 
  until finally dissipated by the viscosity.
Understanding the process of energy cascade 
  has been one of the central problems 
  in fluid turbulence. 

Since the realistic turbulence 
  is difficult to analyze directly,  
  researchers have looked for simpler models of energy cascade.
One of the most successful attempts 
  is known by the name of the shell model
  \cite{Frisch.Book1995,Bohr.Book1998,Biferale.ARFM35}:
\begin{equation}
  \left(\frac{\D}{\D{t}} + \nu k_j^2 \right) u_j 
  = \mathrm{i} k_j \mathcal{Q}_j + F_j^{\mathrm{(ex)}} 
  \quad (j = 1,2,\ldots,N)
  \label{shell},
\end{equation}  
  with the Fourier space discretized into ``shells'' 
  with geometrically increasing wavenumber $k_j$, 
  and the quadratic nonlinear term $\mathcal{Q}_j$
  chosen to conserve the energy
  \cite{Ohkitani.PTP81,Bohr.Book1998,Biferale.ARFM35}.
In the statistically steady turbulence
  maintained by applying the external forcing $F_j^{\mathrm{(ex)}}$ 
  to the shells with small $j$
  so that the energy cascades down to larger wave\-numbers, 
  the model is known to reproduce 
  the essential features of real turbulence, 
  including the Kolmogorov spectrum 
  and the anomalous scaling exponents
  \cite{Frisch.Book1995,Bohr.Book1998,Biferale.ARFM35}.
Even after the supercomputers started to enable  
  direct numerical simulations of turbulent flows, 
  shell models provide a useful platform 
  to test ideas about turbulence
  \cite{Gallavotti.Book2002,Benzi.PRE68,Matsumoto.PRE89}.

Although the advantages of shell models 
  overwhelm its disadvantages \cite{Biferale.ARFM35},
  we have one dissatisfaction:
  the shell model cannot be materialized as a physical apparatus.
This is contrastive to the case 
  of multiple pendulum
  \cite{Oyama.JPS98a,Awrejcewicz.NLD50,Konishi.PRE108}
  whose experimental realizability can help physical intuition.
However, unlike the shell models, 
  conventional multiple pendulums 
  are inconvenient for studies of energy transfer, 
  as every eigen\-mode spreads over the whole pendulum system,
  which makes it difficult to excite a single mode 
  and to define the energy transfer from it.


\begin{figure}
  \centering
  \raisebox{2.1cm}{(a)}\raisebox{-2.0cm}{%
  \IncFig[width=2.5cm]{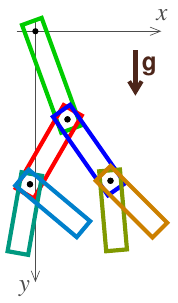}%
  }\quad 
  \parbox{3.6cm}{\centering
  \raisebox{2.0cm}{(b)}\IncFig[width=2.8cm]{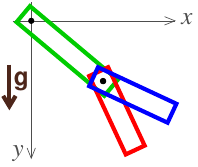}%
  \\
  \raisebox{2.0cm}{(c)}\IncFig[width=2.8cm]{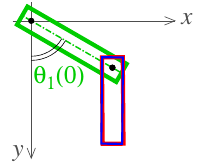}
  }
  \caption{\label{Fig:schem}%
  Schematic description of branched multiple pendulums.
  (a) An example with three generations.
  (b) Branched double pendulum 
  (with ``double'' referring to two generations). 
  (c) Initial condition, in which only the in-phase modes are excited. 
  }
\end{figure}

Considering all these issues, 
  here we propose to study energy transfer 
  in pendulum systems arranged in a tree structure, 
  which we call \emph{branched multiple pendulums}.
Borrowing the words of the family tree, 
  we may describe the system as follows:
  The parent link has two or more child links, 
  each of which can have grandchildren, and so on.
An example of branched triple pendulum 
  is shown in Fig.~\ref{Fig:schem}(a) 
  where ``triple'' refers to three generations.
Obviously 
  the number of generations can be increased systematically. 
For the sake of simplicity,
  the branching number is limited to two,  
  and the links belonging to the same generation 
  are assumed to be equal.

A remarkable feature of branched multiple pendulum
  is the presence of modes localized to younger generations (i.e.\ 
  lower part of the pendulum).
This enables us 
  to study cascading of energy down to descendants. 

In the simplest case shown in Fig.~\ref{Fig:schem}(b), 
  the system consists of three links:
  the parent (link 1)
  and the two children (links 2 and 3).
The linear eigen\-modes 
  split into symmetric and antisymmetric types
  with regard to exchange of the children.
We refer to the former as the in-phase modes
  (with the children oscillating in phase), 
  and the latter as the anti\-phase mode.

The anti\-phase mode 
  is localized to the children. 
This implies 
  that the anti\-phase mode is not excited directly 
  by displacing the parent link alone
  as in Fig.~\ref{Fig:schem}(c).

If the motion starts from the initial condition 
  in which the children are strictly in phase, 
  they remain in phase.
This is reminiscent of the well-known feature 
  of Eq.~(\ref{NS.vort}) 
  that, if the vorticity $\vort$ vanishes initially,
  $\vort$ remains zero even if $\bm{u} \ne \bm{0}$.
If small initial vorticity is present, however, 
  nonlinear interactions in Eq.~(\ref{NS.vort}) may make it grow.
Analogously, 
  anti\-phase oscillation in our pendulum system
  may grow from small initial disturbance.
If such a growth occurs,
  it must be associated with energy transfer 
  from the in-phase modes to the anti\-phase mode.

Here we report 
  that growth of anti\-phase oscillation actually occurs 
  in real experiments 
  starting from the initial condition in Fig.~\ref{Fig:schem}(c), 
  if the initial angle of the parent is sufficiently large.
Then we demonstrate 
  how the energy transfer can be detected in experiment:  
We derive the energy transfer function, denoted with $W$, 
  by calculating the Poisson bracket 
  for the Hamiltonian of the anti\-phase mode, $\Hd$; 
  subsequently, we evaluate $W$ and $\Hd$ 
  with time series data from video image analysis.


\section{Formulation}

\begin{figure}
  \centering
  \raisebox{2.3cm}{(a)}\IncFig[width=3.0cm]{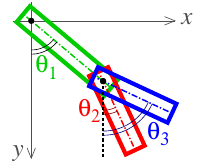}\qquad
  \raisebox{2.3cm}{(b)}\IncFig[width=3.0cm]{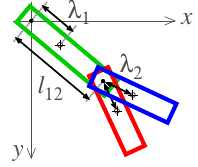}%
  \caption{\label{Fig:setup}%
  Definitions of the angles and lengths 
  for the branched double pendulum. 
  The distance between the pivot and the center of mass 
  is denoted with $\ld_i$ for the link in the $i$-th generation.  
  }
\end{figure}


Taking the angles $(\th_1,\th_2,\th_3)$ 
  shown in Fig.~\ref{Fig:setup}(a)
  as the generalized coordinates, 
  we write the Lagrangian:
\begin{equation}
  L = K - U \relax, \qquad
  K = {\frac12}
  \begin{bmatrix} \dot\th_1 & \dot\th_2 & \dot\th_3 \end{bmatrix}
  \mathsf{M}
  \ThreeVect{\dot\th_1}{\dot\th_2}{\dot\th_3} 
  \label{Lagrangian//th123}, 
\end{equation}
  where the mass matrix $\mathsf{M}$ is given as 
\begin{equation}
  \mathsf{M}
  = 
  \begin{bmatrix}
  M_1  & \mu \cos{\Delta_2} & \mu \cos{\Delta_3} \\
  \mu \cos{\Delta_2} & M_2 & 0   \\
  \mu \cos{\Delta_3} & 0   & M_2 
  \end{bmatrix}
  \label{M//th123}
\end{equation}
  in terms of 
  $\Delta_i = \theta_i - \theta_1$, 
  and the potential $U$ reads 
\begin{equation}
  U = G_1 ( 1 - \cos\th_1 ) +  G_2 ( 2 - \cos\th_2 - \cos\th_3 )
  \label{U//th123}.
\end{equation}
The constants in Eqs.~(\ref{M//th123}) and (\ref{U//th123}) 
  are related to the physical parameters of the pendulum system,   
  including the lengths shown in Fig.~\ref{Fig:setup}(b);
  they are related by 
\begin{gather*}
  M_1 = I_1' + 2 m_2 \Ell^2 \relax, \quad
  M_2 = I_2'                \relax, \quad  
  \mu = m_2 \Ell \ld_2      \relax, \quad \\
  G_1 = m_1 g \ld_1 + 2 m_2 g \Ell \relax, \quad 
  G_2 = m_2 g \ld_2 
  \relax,
\end{gather*} 
  where $m_1$ is the mass of the parent link, 
  $I_1' = I_1 + m_1 \ld_1^2 $ is its moment of inertia 
  about the pivot,
  and $m_2$ and $I_2' = I_2 + m_2 \ld_2^2$ are those of the children.
The gravitational acceleration 
  is denoted with $g$.


Since the two child links are equal,
  the Lagrangian is symmetric 
  with regard to the exchange of $\th_2$ and $\th_3$.
Motivated by this symmetry, 
  we introduce 
\begin{equation}
  \thM = \frac{\th_2 + \th_3}{2} \relax, \quad 
  \thD = \frac{\th_2 - \th_3}{2} \relax.
  \label{thMD}
\end{equation}

By switching from $(\th_1, \th_2, \th_3)$ to $(\th_1, \thM, \thD)$, 
  the Lagrangian in Eq.~(\ref{Lagrangian//th123}) is rewritten as 
\begin{align}
  & L = K - U, \qquad 
  K = {\frac12}
  \begin{bmatrix} \dot\th_1 & \thM[\dot] & \thD[\dot] \end{bmatrix}
  \tilde{\mathsf{M}}
  \ThreeVect{\dot\th_1 }{\thM[\dot]}{\thD[\dot]}
  \label{Lagrangian//thMD},
  \\
  &\tilde{\mathsf{M}} = 
  \begin{bmatrix}
  M_1    &  {2\mu}\cosDeM\cos\thD & {-2\mu}\sinDeM\sin\thD \\
  ~{2\mu}\cosDeM\cos\thD &  2 M_2  &      0  \\  
  {-2\mu}\sinDeM\sin\thD &      0  &  2 M_2
  \end{bmatrix}
  \label{M*},
  \\[1.0ex]%
  &
  U = G_1 ( 1 - \cos\th_1 ) + 2 G_2 ( 1 - \cos\thM \cos\thD )
  \label{U.thMD*}, 
\end{align}
  where we have introduced 
  \[ (\cosDeM,\sinDeM) = (\cos\DeM,\sin\DeM) \] 
  for the sake of brevity
  (with $\DeM = \thM - \th_1$).

Linear analysis for small-amplitude oscillations
  yield two kinds of modes:
  the in-phase oscillation modes, with $\thD=0$, 
  and the anti\-phase oscillation mode, 
  in which $(\th_1,\thM) = (0,0)$ and $\thD\ne0$.
Note that the anti\-phase oscillation 
  is localized to the children.
We denote 
  the eigenfrequencies of the two in-phase modes 
  with $\om_\pm = {2\pi}f_\pm$
  and that of the anti\-phase mode 
  with $\omD = 2\pi\fd$.  


\section{Derivation of energy transfer function}

As theoretical preparation 
  for experimental detection of energy inflow 
  into the anti\-phase mode, 
  here we derive its mathematical expression 
  as a function of the canonical variables.
We begin with transition to the Hamiltonian formulation
  and extraction of the sub-Hamiltonian for the anti\-phase oscillation.
  
Using $\tilde{\mathsf{M}}$ in Eq.~(\ref{M*}),
  we have 
\begin{equation}
  \ThreeVect{p_1}{\pM}{\pD}
  = \tilde{\mathsf{M}}\ThreeVect{\dot\th_1}{\thM[\dot]}{\thD[\dot]}
  \label{pMD//thMD.dot}
\end{equation}
  as the generalized momenta. 
The total Hamiltonian reads 
\begin{equation}
  H = K + U,  \quad 
  K = 
  {\frac12} 
  \begin{bmatrix} p_1 & \pM & \pD \end{bmatrix}
  \tilde{\mathsf{M}}^{-1}
  \ThreeVect{p_1}{\pM}{\pD}
  \label{H},
\end{equation}
  from which we will extract 
  the sub-Hamiltonian, $\Hd = \Hd(\thD,\pD)$, 
  independent of the in-phase mode variables 
  $(\th_1, p_1,\thM, \pM)$.

In extracting the kinetic energy of the anti\-phase mode from $K$, 
  we make use of the feature of $\tilde{\mathsf{M}}$
  that it becomes block\-wise diagonalized
  for small angles.
This suggests isolating $\pD^2/(4M_2)$ 
  as a term including the anti\-phase mode variables alone.

For the potential energy, 
  though we could extract $G_1 \thD^2$
  assuming small angles,
  here we employ a more refined expression 
  taking the ambiguity of the angles into account, 
  in view of future application to finite angles. 
Since $\thD$ has ambiguity of $\pi$ 
  due to the $2\pi$ indeterminacy of $\th_2$ and $\th_3$, 
  the extracted potential energy should be invariant 
  under the shift of $\thD$ by $\pm\pi$.
This requirement is met by employing 
  $G_1 {\sin^2}\thD$ instead $G_1 \thD^2$.

Thus we have 
\begin{equation}
  \Hd = \frac{\pD^2}{4M_2} + G_2 \sin^2\thD 
  \label{Hd}
\end{equation}
  as the sub-Hamiltonian of the anti\-phase mode.


With $\Hd$ given in Eq.~(\ref{Hd}),
  the Poisson bracket between $\Hd$ and $\Hm = H - \Hd$
  gives the energy transfer function, 
\begin{align}
  W = \dd{\Hd}{\thD} \dd{\Hm}{\pD} -  \dd{\Hd}{\pD} \dd{\Hm}{\thD}
  \label{W//Hd.poisson},  
\end{align}
  such that $\D{\Hd}/\D{t} = W$.
In principle,
  $W$ can be explicitly obtained 
  as a function of the canonical variables.

It turns out to be difficult, in practice, 
  to deal with the inverse matrix $\tilde{\mathsf{M}}^{-1}$ in $\Hm$
  originating from Eq.~(\ref{H}).
Here we simplify the calculation 
  by assuming $\abs{\thD}\ll1$, 
  knowing 
  that $\tilde{\mathsf{M}}$ becomes block\-wise diagonal 
  in the limit of $\thD\to0$. 
For small but finite $\thD$, 
  to a first-order approximation, 
  we have
\begin{multline}
  \tilde{\mathsf{M}}^{-1} 
  = 
  \begin{bmatrix}
  \dfrac{2M_2}{D} & {-\dfrac{2\mu\cosDeM}{D}} & 0 \\[2.0ex]
  {-\dfrac{2\mu\cosDeM}{D}} & \dfrac{M_1}{D}  & 0 \\[1.0ex]
  0 & 0 & \dfrac{1}{2M_2} 
  \end{bmatrix}
  \\ {}
  + \frac{2\mu\thD\sinDeM}{D} 
  \begin{bmatrix}
  0 & 0 & 1 \\
  0 & 0 & {-\dfrac{\mu\cosDeM}{M_2}}      \\ 
  1 &     {-\dfrac{\mu\cosDeM}{M_2}} & 0   
  \end{bmatrix}
  + O(\thD^2)
  \label{invM*}
\end{multline}
  where $D = 2 M_1 M_2 - 4\mu^2 \cos^2\DeM $. 
By substituting Eq.~(\ref{invM*}) 
  into $\tilde{\mathsf{M}}^{-1}$ in Eq.~(\ref{H}), 
  we obtain
\begin{multline}
  K
  = \frac{M_2}{D} p_1^2 
  + \frac{( M_1\pM - {4\mu\cosDeM} p_1 )\pM}{2D}
  + \frac{\pD^2}{4M_2}     
   \\ {} 
  + \frac{2\mu\sinDeM ( M_2 p_1 - {\mu\cosDeM}\pM)  \pD\thD}{M_2 D}
  + O(p^2\thD^2)
  \label{K.approx}.
\end{multline}
Combining Eq.~(\ref{K.approx}) 
  with Eq.~(\ref{U.thMD*}) for $U$, 
  we write down $\Hm = K + U - \Hd$ explicitly, 
  from which the Poisson bracket
  on the right-hand side of Eq.~(\ref{W//Hd.poisson}) 
  is calculated. 
As a result, we have
\begin{subequations}%
\begin{align}
  W &= \WK + \WP  \label{W}
\intertext{where}
  \WK &= \frac{%
  (\mu\sin\DeM) \{ -M_2 p_1 + (\mu\cos\DeM)\pM \} \pD^2}{M_2^2 D}
  \label{WK}, \\
  \WP &= \frac{G_2}{M_2} (1-\cos\thM) \pD\thD
  \label{WP}.
\end{align}%
\label{eqs:W}%
\end{subequations}%


\section{Experimental procedure}

We performed experiments 
  on a pendulum system consisting of three links 
  made of transparent acrylic plates. 
To each link we attached two color dot stickers,
  whose positions were record with a camera (Canon EOS Kiss M)
  and tracked by means of video image analysis.


The lengths of the links, defined in Fig.~\ref{Fig:setup}(b), 
  are
\begin{equation}
  \ld_1 = 8.4\,\cm, \quad 
  \ld_2 = 4.3\,\cm, \quad  \ell_{12} = 27.0\,\cm ; 
\end{equation}
  and their masses were measured directly: 
\begin{equation}  
  m_1 = 0.12880\,\kg \relax, \quad 
  m_2 = 0.10531\,\kg \relax. 
\end{equation}
The remaining constant parameters in the Lagrangian, 
  namely the moments of inertia, $I_1$ and $I_2$, 
  were evaluated through the eigenfrequencies.
From video image analysis data of small-amplitude oscillations
  started from several different initial conditions, 
  with FFT we obtained  
\begin{equation}
  {f_+ = 0.88\,\Hz},\quad  {f_- = 1.58\,\Hz}, \quad
  {\fd = 1.26\,\Hz}
  \label{SysIdent.freq}\relax. 
\end{equation}%
By matching the theoretical eigenfrequencies 
  with the experimental values in Eq.~(\ref{SysIdent.freq}), 
  we determined the moments of inertia as
\begin{equation}
  I_1 = 1.88  \times 10^{-3}\,\kg\,\mathrm{m}^2 \relax, \quad 
  I_2 = 0.516 \times 10^{-3}\,\kg\,\mathrm{m}^2 \relax 
  \label{SysIdent.I}. 
\end{equation}%


Energy transfer experiments were then performed 
  under the initial condition illustrated in Fig.~\ref{Fig:schem}(c).
We displaced the parent link 
  by the intended initial angle $\th_1(0)$, 
  suspended it with a nylon thread,  
  and waited until the child links come to rest 
  in the vertical position.
Then we released the pendulum by cutting the thread
  at the time $t = t_0$.
Each video record was started a few seconds before $t_0$,
  so that the actual value of the initial angle
  can be obtained from the image data.

\section{Results}

\begin{figure}
  \centering
  \raisebox{3.0cm}{(a)}%
  \IncFig[width=7.0cm]{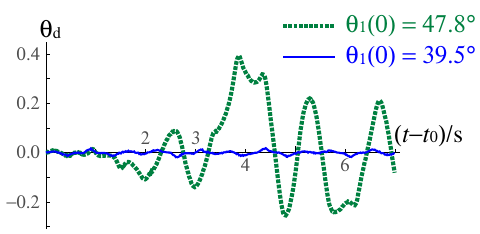}\\[0.5ex]%
  \raisebox{3.3cm}{(b)}%
  \IncFig[width=7.0cm]{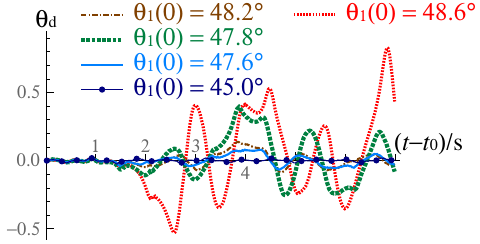}
  \caption{\label{Fig:thD:395-486}%
    Behavior of the anti\-phase mode variable $\thD$ 
    for different values of the initial parent angle $\th_1(0)$.
    (a) Comparison of the cases 
    with $\th_1(0) = 39.5\Deg$ and $\th_1(0) = 47.8\Deg$.
    Growth of $\thD$ is observed in the latter case.
    (b) Comparison of cases with five different values of $\th_1(0)$, 
    ranging from $45.0\Deg$ to $48.6\Deg$.
  }
\end{figure}

As a result of these experiments, 
  the anti\-phase mode 
  was found to grow out of $\thD(0) \approx 0$ 
  into appreciable oscillations, 
  if $\th_1(0)$ is large enough.
Behavior of $\thD$  
  is shown in Fig.~\ref{Fig:thD:395-486}.
The growth depends on the largeness of $\th_1(0)$,  
  as is evident from Fig.~\ref{Fig:thD:395-486}(a)
  in which two cases are compared: 
In the case of $\th_1(0) = 39.5\Deg$,
  the oscillation amplitude of $\thD$ remains 
  as small as given initially, 
  while the amplitude of $\thD$ 
  for a larger initial value of the parent angle, 
  $\th_1(0) = 47.8\Deg$, 
  grows many times greater than the initial disturbance.

We repeated the experiment 
  with about twenty different values of $\th_1(0)$.
For $\th_1(0) \le 47.6\Deg$,  
  no significant growth of $\thD$ was observed.
The anti\-phase mode was excited 
  basically for $\th_1(0) \ge 47.8\Deg$, 
  as is shown in Fig.~\ref{Fig:thD:395-486}(b).

\begin{figure}
  \centering
  \IncFig[width=7.5cm]{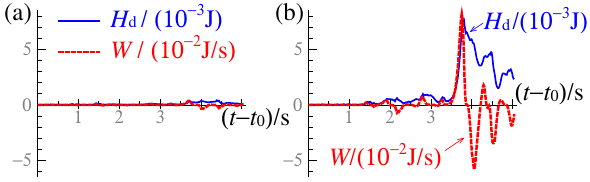}%
  \caption{\label{Fig:W:476-478}%
  $\Hd$ and $W$ for (a) $\theta_1(0)=47.6\Deg$
  and (b) $\theta_1(0)=47.8\Deg$\relax.}
\end{figure}


Having known 
  that $\thD$ grows significantly for $\th_1(0) = 47.8\Deg$ 
  but not for $\th_1(0) = 47.6\Deg$, 
  we examined the energy transfer in these two cases.
We evaluated $\Hd$ and $W$ 
  according to Eqs.~(\ref{Hd}) and (\ref{eqs:W}),  
  using the time series data of $\th_i(t)$ from video image analysis.
The results 
  are shown in Fig.~\ref{Fig:W:476-478}. 
Note that the vertical scales are unified 
  for proper comparison.

In the case of $\th_1(0) = 47.6\Deg$ in Fig.~\ref{Fig:W:476-478}(a),  
  the values of $\Hd$ and $W$ are almost vanishing,
  showing only small fluctuations. 
Contrastively, 
  $\Hd$ grows considerably
  in the case of $\th_1(0) = 47.8\Deg$,
  as is seen in Fig.~\ref{Fig:W:476-478}(b). 
This growth 
  is associated with inflow of energy into the anti\-phase mode.
After the pendulum is released at $t=t_0$,
  the energy transfer function $W$ 
  departs from zero and remains mostly positive, 
  until $\Hd$ reaches its maximum at $t - t_0 = 3.8\,\mathrm{s}$. 
Thus influx of energy into the anti\-phase mode  
  is detected.
It should be noted, however, 
  that the behavior of $W$ for later time after the growth of $\thD$ 
  can be questionable, 
  as the approximation of small $\thD$ in Eq.~(\ref{K.approx}) 
  becomes inaccurate.

\begin{figure}
  \centering
  \IncFig[width=7.0cm]{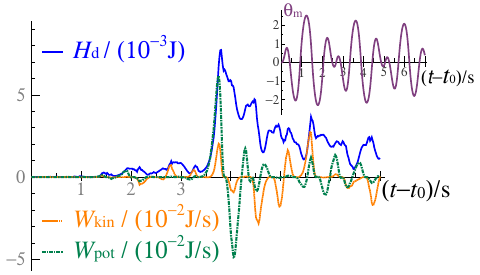}%
  \caption{\label{Fig:W.detail}%
    $\WK$ and $\WP$ from the same data 
    as in Fig.~\protect\ref{Fig:W:476-478}(b), 
    plotted along with $\Hd$. 
    Inset: $\thM$ from the same data.
  }
\end{figure}

The two components of $W$ in Eq.~(\ref{W}), 
  namely $\WK$ and $\WP$,  
  are evaluated separately in Fig.~\ref{Fig:W.detail}.
The contribution of $\WP$ is dominant,  
  although not overwhelming.
The large contribution of $\WP$ 
  is consistent with largeness of $\thM$, 
  whose maximum exceeds 2 radians.


\begin{figure}
  \centering
  \IncFig[width=6.5cm]{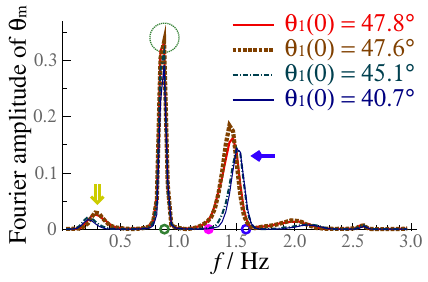}%
  \caption{\label{Fig:qM.Fourier}%
  Fourier spectrums of $\thM$ 
  for $0 \le {t-t_0} < 32\,\mathrm{s}$.
  The three linear eigenfrequencies in Eq.~(\protect\ref{SysIdent.freq}),
  namely $f_\pm$ and $\fd$, are marked on the $f$-axis.
  }
\end{figure}

Noticing that $\WP$ in Eq.~(\ref{WP}) 
  is essentially the product of $1-\cos\thM$ and $\pD\thD$, 
  with the latter oscillating with the frequency $2\fd$, 
  we carried out Fourier analysis of $\thM$.  
Combination of modes in $\thM$
  can make $W$ positive, 
  if their frequencies sums up to $2\fd$.

The Fourier spectrums of $\thM$
  are shown in Fig.~\ref{Fig:qM.Fourier}, 
  along with the linear eigenfrequencies marked on the $f$-axis.
The eigenfrequencies are non-resonant, 
  as the sum of the two in-phase frequencies, 
  $f_+ + f_- = 2.46\,\Hz$,  
  is smaller than $2\fd = 2.52\,\Hz$.
The nonlinear effect of the finite amplitude
  on the Fourier spectrum is twofold:
While the peak corresponding to $f_-$ 
  shifts toward lower frequencies, 
  making the two in-phase modes 
  even more distant from resonance with $2\fd$, 
  the nonlinear interactions also produce 
  low-frequency components around $0.3\,\Hz$.
It is plausible 
  that these low-frequency components
  fill the mismatch 
  between the two in-phase mode frequencies 
  and twice the anti\-phase mode frequency.


\section{Concluding remarks}

To summarize, 
  in search of models
  allowing dynamical system approach to turbulence 
  and materializable as a physical apparatus, 
  we proposed to study 
  energy transfer in branched multiple pendulums. 
As a first step, 
  we analyzed the energy transfer function $W$ 
  in a branched double pendulum.
By experiments under initial conditions 
  with $\th_1(0) > 0$ and $(\th_2(0),\th_3(0))\approx(0,0)$, 
  we observed that $\thD$ grows out of $\thD(0)\approx 0$
  for $\th_1(0) \ge 47.8\Deg$. 
Energy transfer into $\Hd$  
  was detected by evaluating $W$.

The largeness of the critical initial angle $\th_1(0)$, 
  almost reaching one radian, 
  is probably not peculiar to the present setup
  but universal to a broader range of conditions.
Large values of $\th_1(0)$ are needed for growth of $\thD$ 
  also in another apparatus \cite{Toda.JPS25a}, 
  as well as in most of preliminary numerical simulations
  \cite{Miura+Kato.thesis}. 
Systematic development of numerical and theoretical studies 
  on the critical value of $\th_1(0)$
  is now planned and will be reported elsewhere.

As an obvious future direction, 
  we may increase the generations in the pendulum system 
  and add dissipation to the dynamics of the youngest generation, 
  making the model closer toward the fluid turbulence.
It is worthwhile to investigate, for example,
  under what conditions 
  the Kolmogorov spectrum \cite{Frisch.Book1995} 
  can be reproduced with the pendulum model.
The tree-like structure of branched pendulums 
  may be also relatable 
  to wavelet expansion approach to turbulence \cite{Nakano.PhF31}.
We hope that the present work offers a valuable starting point 
  for these possibilities.

\section*{Acknowledgments}

The authors gratefully acknowledge valuable comments 
  from Drs.\ 
  Mikito Toda, Kazuyuki Yoshimura, 
  So Kitsune\-zaki, Shin\-taro Nakatani, 
  Masaru Furukawa, Toshiyuki Doi, Tomonari Nakayama, 
  Tsuyoshi Chawan\-ya,  
  Yasushi Shimizu,  Hiroyasu Katsuno,  Hiroshi Takano,  
  Takeshi Matsu\-moto, Susumu Goto, and Michio Otsuki.
English language polishing 
  was partially assisted by Google Gemini.
This work was supported 
  by JSPS \textsc{Kaken\-hi} Grant Number JP-24K06887.


\begingroup
\providecommand{\newblock}{\relax}
\bibliographystyle{jpsj} 
\bibliography{turbulence,statmech,Book,ref2605,thesis26e}
\endgroup
\vfill
\end{document}